\documentclass[prl,twocolumn,floats,aps,epsfig,nofootinbib,amssymb]{revtex4}
\usepackage{subfigure}
\def\beq{\begin{equation}}
\def\eeq{\end{equation}}
\def\bea{\begin{eqnarray}}
\def\eea{\end{eqnarray}}

\def\dl{\Delta L}

\def\dl{\Delta L}
\def\vd{v_\Delta^{}}
\def\md{M_\Delta^{}}

\def\be{\begin{equation}}
\def\ee{\end{equation}}
\def\bea{\begin{eqnarray}}
\def\eea{\end{eqnarray}}

\def\pslash{\not{\hbox{\kern-4pt $p$}}}
\def\qslash{\not{\hbox{\kern-4pt $q$}}}
\def\lv{\not{\hbox{\kern-4pt $L$}}}
\def\lsim{\mathrel{\raise.3ex\hbox{$<$\kern-.75em\lower1ex\hbox{$\sim$}}}}
\def\gsim{\mathrel{\raise.3ex\hbox{$>$\kern-.75em\lower1ex\hbox{$\sim$}}}}
\def\ifmath#1{\relax\ifmmode #1\else $#1$\fi}

\usepackage{graphicx}
\usepackage{bm}
\begin{document}
\draft
\renewcommand{\thefootnote}{\arabic{footnote}}

\title{Testing a Neutrino Mass Generation Mechanism at the LHC}
%\bigskip
\author{Pavel Fileviez P{\'e}rez$^1$, Tao Han$^{1,2}$, Gui-Yu Huang$^1$,
Tong Li$^{1,3}$, Kai Wang$^1$}
\address{
$^1$Department of Physics, University of Wisconsin, Madison, WI 53706, USA\\
$^2$KITP, University of California, Santa Barbara, CA 93107, USA\\
$^3$Department of Physics, Nankai University, Tianjin 300071, P.R.~China }
\date{\today}
\begin{abstract}
The Large Hadron Collider could be a discovery machine for the
neutrino mass pattern and its Majorana nature in the context of a
well-motivated TeV scale Type II seesaw model. This is achieved by
identifying the flavor structure of the lepton number violating
decays of the charged Higgs bosons. The observation of either
$H^+ \to \tau^+\bar \nu$ or $H^+ \to e^+\bar \nu$ will be
particularly robust to determine the neutrino spectra since they are
independent of the unknown Majorana phases, which could be probed via
the $H^{++} \to e^+_i e^+_j$ decays. In a less favorable scenario when the
leptonic channels are suppressed, one needs to observe the
decays $H^+ \to W^+ H_1$, and $H^+ \to t\bar b$ to confirm the
triplet-doublet mixing that implies the Type II relation.
The associated production $H^{\pm \pm} H^{\mp}$ is crucial in order
to test the triplet nature of the Higgs field.
\end{abstract}
\maketitle
\noindent
{\bf{I. Introduction}}\\
The Large Hadron Collider (LHC) at CERN will soon take us to a new
frontier with unprecedented high energy and luminosity. Major discoveries
of exciting new physics at the Terascale are highly anticipated. The
existence of massive neutrinos clearly indicates the need for new physics
beyond the Standard Model (SM)~\cite{review}. It is thus pressing to
investigate the physics potential of the LHC in this regard.
The leading operator relevant for neutrino masses in the context
of the SM \cite{Weinberg} is $( \kappa /\Lambda) l_L^{} l_L^{} HH$,
where $l_L^{}$ and $H$ stand for the $SU(2)_L$ leptonic and Higgs doublet,
respectively. After the electroweak symmetry breaking (EWSB), the neutrino
Majorana mass reads as $m_\nu \sim \kappa v^2_0/\Lambda$, where
$v_0 \approx 246$ GeV. The crucial issue is to understand the origin of
this operator in a given extension of the SM in order to identify the
dimensionless coupling $\kappa$ and the mass scale $\Lambda$ at which
the new physics enters.

There exist several simple renormalizable extensions of the SM to generate
neutrino Majorana masses and mixing. (i) The simplest one is perhaps
the Type-I seesaw mechanism~\cite{TypeI}, where one adds fermionic gauge
singlets $N$. The resulting neutrino mass is given by
$m_\nu \propto v^2_0/ M_N$. The smallness of $m_\nu \lsim 1$ eV is thus
understood by the ``seesaw" spirit if $M_N \gg v_0$. The interests of
searching for heavy Majorana neutrinos $N$ at the LHC have been lately
renewed~\cite{Han:2006ip}. However, it is believed that any signal of
$N$ would indicate a more subtle mechanism beyond the simple Type I seesaw
due to the naturally small mixing $V_{N\nu}^2 \sim m_\nu/M_N$. (ii) A more
appealing mechanism, at least from the phenomenological point of view,
is the Type II seesaw mechanism~\cite{TypeII}. In this scenario the Higgs
sector of the SM is extended by adding an $SU(2)_L$ Higgs triplet,
$\Delta \sim (1,3,1)$ under the SM gauge groups. After EWSB the neutrino mass
is given by $m_\nu \propto Y_\nu v^{}_{\Delta}$, where $Y_\nu$ and
$v_{\Delta}$ are the Yukawa coupling and the vacuum expectation value of
the triplet, respectively. If the triplet mass is accessible at the
LHC, $\md \lesssim {\cal O}(1)$ TeV, then this scenario may lead to
very rich phenomenology. Experimentally verifying this mechanism would
be of fundamental importance to understand the neutrino mass generation
and its connection to the EWSB. For other proposals with exotic leptonic
representations or radiative mass generations, see~\cite{TypeIII}.

In this Letter, we explore the feasibility to test the Type II seesaw
mechanism at the LHC assuming that the Higgs triplet is kinematically
accessible. We focus on the exciting possibility to determinate the neutrino
spectrum through the lepton violating Higgs decays in the theory.
Recently, several groups~\cite{Thomas, Raidal, Akeroyd} have studied
the possibility to distinguish between the neutrino spectra using
the predictions for the decays of the doubly charged Higgs.
Unfortunately, this method suffers from the dependence on
the unknown Majorana phases.
%can be used only when the lightest neutrino
%mass is much smaller than $10^{-2}$ eV, since these decays are quite
%dependent of the unknown Majorana phases.
Here we point out
for the first time that the best way to determinate the neutrino spectrum
in this context is through the lepton-number violating decays
$H^+ \to e^+_i \bar{\nu}\ (i=e,\mu,\tau)$,
since those are independent of the unknown Majorana phases. We advocate that the
associated production $H^{\pm \pm} H^{\mp}$ is crucial in order to test the triplet nature
of the model. With semi-realistic Monte Carlo simulations,
we demonstrate how to reconstruct the signal events
$H^{\pm \pm} H^{\mp} \to e^{\pm}_i e^{\pm}_j e^{\mp}_k \nu$
and suppress the backgrounds up to 1 TeV of the Higgs mass.
We also show how to test this theory when
the leptonic channels are suppressed. The discovery of the predicted signals
at the LHC would provide us crucial information about the neutrino
mass and its connection to the electroweak symmetry breaking mechanism.
%%%%%%%%%%%%%%%%%%%%%%%%%%%%%%%%%%%%%%
\vskip 0.3mm
\noindent
{\bf II. The Type II Seesaw for Neutrino Masses}\\
The Higgs sector of the Type II seesaw scenario is composed of
the SM Higgs $H (1,2,1/2)$ and a scalar triplet $\Delta (1,3,1)$.
The crucial terms for the neutrino mass generation in the theory are
\begin{eqnarray}
 - Y_\nu \ l_L^T \ C \ i \sigma_2 \ \Delta \ l_L  +
  \mu \ H^T \ i \sigma_2 \ \Delta^\dagger H \ + \ \text{h.c.}
\label{Yukawa}
\end{eqnarray}
where the Yukawa coupling $Y_\nu$ is a $3\times 3$ complex symmetric matrix.
The lepton number is explicitly broken by two units due to the simultaneous
presence of the Yukawa coupling $Y_\nu$ and
the Higgs term proportional to the $\mu$ parameter. From the minimization
of the scalar potential one finds $v_{\Delta}=\mu v_0^2/\sqrt{2} M_{\Delta}^2$.
Therefore, the neutrinos acquire a Majorana mass given by
\begin{eqnarray}
M_{\nu}= \sqrt{2} \ Y_\nu \ v_{\Delta} = Y_\nu\  {\mu \ v_0^2}/{  M_{\Delta}^2}.
\label{type2}
\end{eqnarray}
This equation is the key relation of the Type II seesaw scenario.
The neutrino mass is triggered by the EWSB and its smallness is associated
with a large mass scale $\md$. With appropriate choices of the Yukawa matrix elements,
one can easily accommodate the neutrino masses and mixing consistent with the
experimental observation. For the purpose of illustration, we adopt the values of the
masses and mixing at $2\sigma$ level from a recent global fit\cite{Schwetz}.
%%%%%%%%%%%%%%%%%%%%%%%%%%%%%%%%%%%%
\vskip 0.3mm
\noindent
II.A {\textit{ General Properties of the Higgs Sector}}\\
After the EWSB, there are seven massive physical Higgs bosons:
$H_1,\ H_2,\ A, \  H^{\pm}$, and $H^{\pm\pm}$, where
$H_1$ is SM-like and the rest of the Higgs states are $\Delta$-like.
Neglecting the Higgs quartic interactions one finds
$M_{H_2}\simeq M_{A} \simeq M_{H^+} \simeq M_{H^{++}}=M_{\Delta}$.
Since we are interested in a mass scale accessible at the LHC,
we thus focus on $110~{\rm GeV} < \md <   1~{\rm TeV}$,
where the lower bound is from direct searches~\cite{Tevatron}. Working
in the physical basis for the fermions we find that
the Yukawa interactions can be written as
\begin{eqnarray}
\nonumber
& & \nu_L^T \ C \ \Gamma_+ \ H^+ \ e_L, \ \ \text{and}
\ \  e_L^T \ C \ \Gamma_{++} \ H^{++} \ e_L, \\
&& \Gamma_+  = \frac{ c_{\theta_+} m_\nu^{diag} V_{PMNS}^\dagger}{v_{\Delta}}, \
\Gamma_{++} =  \frac{V_{PMNS}^* m_{\nu}^{diag} V_{PMNS}^{\dagger}}{\sqrt{2} \ v_{\Delta}},
\nonumber
\end{eqnarray}
where $c_{\theta_+}= \cos \theta_+$, $\theta_+$ is the mixing
angle in the charged Higgs sector and $v_{\Delta} \lesssim 1$ GeV
from the $\rho$-parameter constraints.
$V_{PMNS}=V_{l} (\theta_{12}, \theta_{23}, \theta_{13}, \delta) \times K_M$
is the leptonic mixing matrix and $K_M=\text{diag} (e^{i\Phi_1/2}, 1, e^{i \Phi_2/2})$
is the Majorana phase factor. The values of the physical couplings
$\Gamma_+$ and $\Gamma_{++}$ are thus governed by the spectrum and mixing
angles of the neutrinos, and they in turn characterize the branching
fractions of the $\dl=2$ Higgs decays. For a previous study of the
doubly charged Higgs decays see~\cite{Chun:2003ej}.

The two leading decay modes for the heavy Higgs bosons are
the $\dl=2$ leptonic mode and the (longitudinal) gauge boson
pair mode. The ratio between them for the $H^{++}$ decay reads as
\begin{eqnarray}
{\Gamma(H^{++}\to \ell^+\ell^+) \over \Gamma(H^{++}\to W^+W^+) }
\approx { | \Gamma_{++} |^2 v_0^4 \over M_\Delta^2   v_\Delta^2}
\approx \left( { m_\nu \over M_\Delta^{}  } \right)^2
 \left({v_0 \over v_\Delta}\right)^4,\
% \nonumber
\end{eqnarray}
using $m_\nu/\md\sim$ 1 eV/1 TeV, one finds that these two decay
modes are comparable when $v_\Delta^{} \approx 10^{-4}\ {\rm GeV}$.
It is thus clear that for a smaller value of $\vd$ (a larger Yukawa coupling),
the leptonic modes dominate, while for larger values, the gauge boson
modes take over. In the case of the singly charged Higgs, $H^{\pm}$,
there is one additional mode to a heavy  quark pair. The ratio
between the relevant channels is
\begin{eqnarray}
{\Gamma(H^{+}\to t \bar b) \over \Gamma(H^{+}\to W^+Z) }
\approx {3 (\vd m_t/v_0^2)^2 \md \over M_\Delta^3   v_\Delta^2 /2 v_0^4 } =
6 \left( { m_t \over M_\Delta^{}  } \right)^2.~~
%\nonumber
\end{eqnarray}
Therefore, the decays $H^+ \to W^+ Z,\ W^+ H_1$ dominate over $t\bar b$
for $\md > 400$ GeV. We present a more detailed discussion elsewhere~\cite{big}.
In our discussions thus far, we have assumed the mass degeneracy for the Higgs triplet.
Even if there is no tree-level mass difference, the SM gauge interactions generate the
splitting of the masses via radiative corrections, leading to
$\Delta M=M_{H^{++}}-M_{H^{+}}=540~\text{MeV}$~\cite{strumia}.
The transitions between two heavy triplet Higgs bosons via the SM gauge
interactions, such as the three-body decays
$H^{++} \to H^+ W^{+*},\  H^{+} \to H^0 W^{+*}$
may be sizable if kinematically accessible. We find~\cite{big} that
these transitions will not have a significant branching
ratio unless $\Delta M > 1 $ GeV. In fact, our analyses will remain valid
as long as $H^{++}$ and $H^+$ are the lower-lying states in the triplet and they are nearly
degenerate. We will thus ignore the mass-splitting effect in the current study.
\begin{figure}[tb]
\includegraphics[scale=1,width=5.0cm]{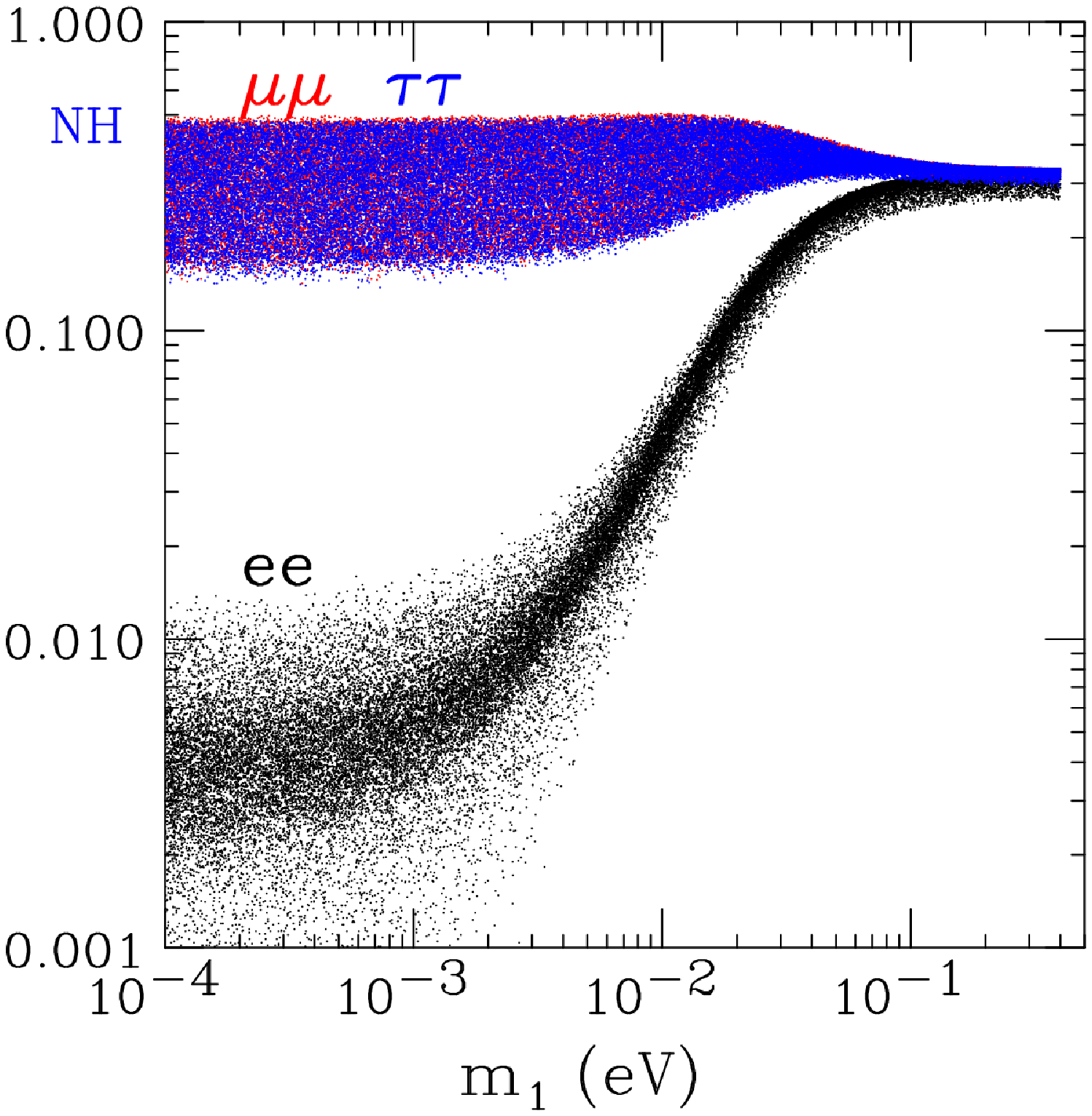}
\includegraphics[scale=1,width=5.0cm]{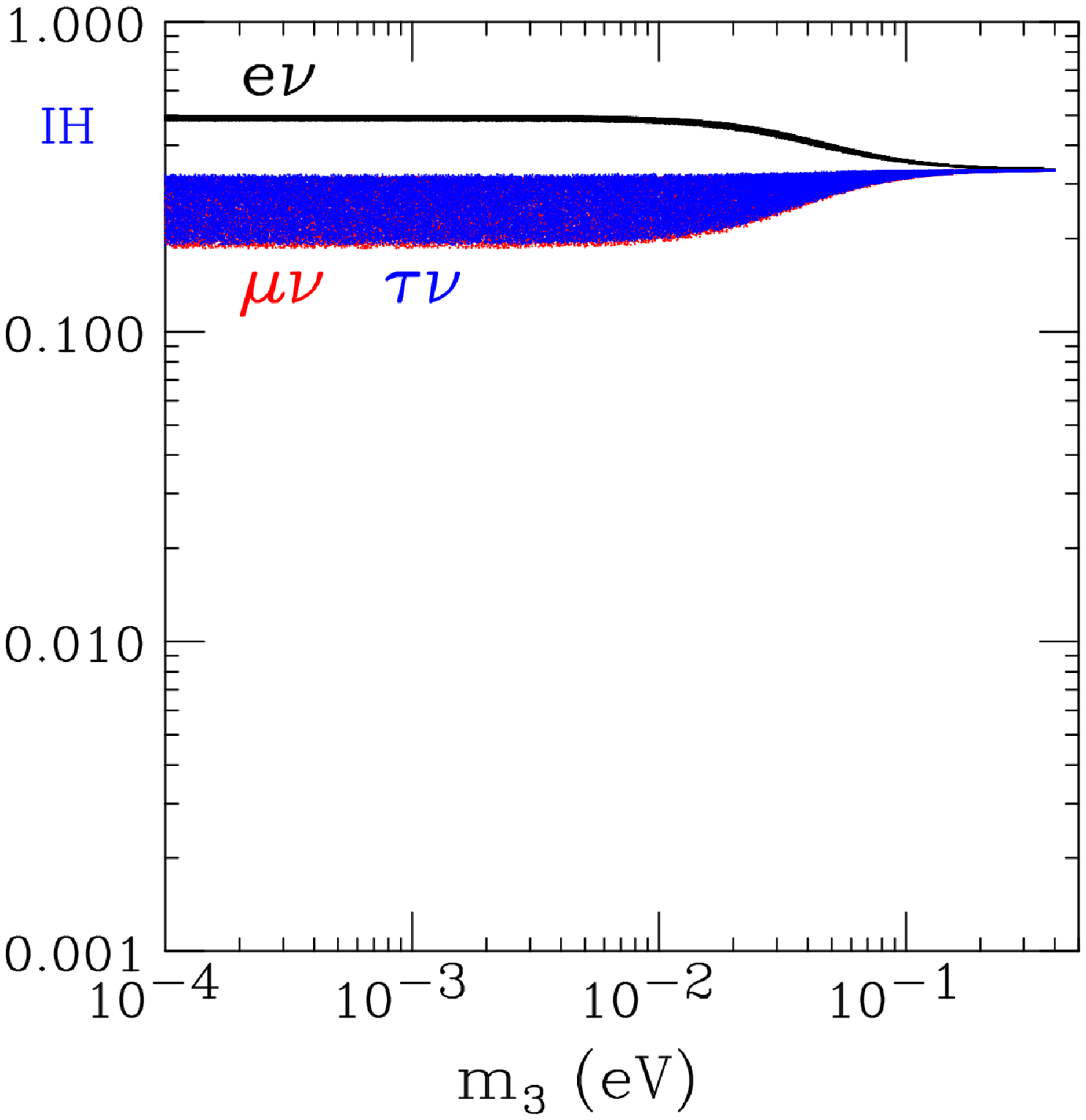}
\caption{\small{Leptonic branching fractions vs
the lightest neutrino mass when $\Phi_i=0$.
(a) for $H^{++}$ decay in the NH, and  (b) for $H^{+}$ in the IH. } }
\label{brii-NH}
\end{figure}

\noindent
II.B {\textit{Higgs Decays and the Neutrino Properties}}
\\
For $v_\Delta^{} < 10^{-4}$ GeV, the dominant channels for the heavy Higgs
boson decay are the $\dl=2$ di-leptons. In Fig.~\ref{brii-NH} we show the
predictions for the representative decay branching fractions (BR) to
flavor diagonal di-leptons versus the lightest neutrino mass where
the spread in BR values is due to the current errors in the neutrino
masses and mixing. Fig.~\ref{brii-NH}(a) is for the $H^{++}$ decay to same-sign di-leptons
in the Normal Hierarchy (NH) ($\Delta m_{31}^2 >0$), and Fig.~\ref{brii-NH}(b)
for the $H^{+}$ decay in the Inverted Hierarchy (IH) ($\Delta m_{31}^2 < 0$).
In accordance with the NH spectrum and the large atmosphere mixing ($\theta_{23}$),
the leading channels are $H^{++}\to \tau^+ \tau^+,\ \mu^+ \mu^+$,
and the channel $e^+ e^+$ is much smaller. When the spectrum is inverted,
the dominant channel is $H^{++} \to e^+ e^+$ instead.
Also is seen in Fig.~\ref{brii-NH}(b) the $H^{+}\to e^+\bar\nu$ dominance
in the IH. In the case of NH the dominant channels are $H^{+}\to \mu^+\bar\nu$
and $H^{+}\to \tau^+\bar\nu$. In both cases of NH and IH, the off-diagonal channel
$H^{++} \to \tau^+ \mu^+$ is dominant due to the large mixing.
In the limit of Quasi-Degenerate (QD) neutrinos one
finds that the three diagonal channels are quite similar,
but the off-diagonal channels are suppressed.

\begin{figure}[tb]
\includegraphics[scale=1,width=5.0cm]{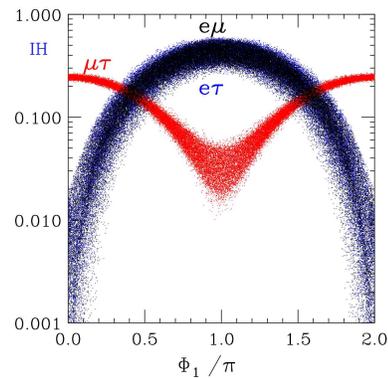}
\caption{\small{Leptonic branching fractions of
$H^{++}$ decay versus the Majorana phase $\Phi_1$
 in the IH for $m_3 \approx 0$.}}
\label{Majorana2}
\end{figure}

The properties of all leptonic decays of the charged
Higgs bosons are summarized in Table.~\ref{Tab1}.
\begin{table}[tb]
\caption{\label{contributions}
{\small{Relations  for the $\dl=2$ decays of $H^{++},\ H^+$ in three different
neutrino mass patterns when $\Phi_1= \Phi_2= 0$.}}}
\begin{ruledtabular}
\begin{tabular}{lcc}
\text{Spectrum} & Relations \\
\hline
NH   & Br$(\tau^+ \tau^+ )$, Br$(\mu^+ \mu^+) \gg$ Br$(e^+ e^+ )$ \\
$\Delta m_{31}^2 > 0$  & Br$(\mu^+ \tau^+) \gg $ Br$(e^+ \tau^+)$, Br$(e^+ \mu^+)$\\
     & Br$(\tau^+ \bar{\nu})$, Br$(\mu^+ \bar{\nu}) \gg $ Br$(e^+ \bar{\nu})$ \\
\hline
IH & Br$(e^+ e^+) > $ Br$(\mu^+ \mu^+)$, Br$(\tau^+ \tau^+)$\\
$\Delta m_{31}^2 < 0$ & Br$(\mu^+ \tau^+) \gg $ Br$(e^+ \tau^+)$,  Br$(e^+ \mu^+)$\\
 & Br$(e^+ \bar{\nu}) > $ Br$(\mu^+ \bar{\nu})$, Br$(\tau^+ \bar{\nu})$ \\
 \hline
QD & Br$(e^+ e^+) \approx$ Br$(\mu^+ \mu^+) \approx$ Br$(\tau^+ \tau^+)$\\
 & Br$(\mu^+ \tau^+) \approx$ Br$(e^+ \tau^+) \approx$ Br$(e^+ \mu^+)$ (suppressed)\\
 & Br$(e^+ \bar{\nu}) \approx$ Br$(\mu^+ \bar{\nu}) \approx$ Br$(\tau^+ \bar{\nu})$
\end{tabular}
\end{ruledtabular}
\label{Tab1}
\end{table}
The effects of the Majorana phases are neglected so far.
The Higgs decays are not very sensitive to the phase $\Phi_2$,
with a maximal reduction of $H^{++}\to \tau^+\tau^+, \mu^+\mu^+$
and enhancement of $\mu^+\tau^+$ up to a factor of two in the NH.
The phase  $\Phi_1$, however,  has a dramatic impact on the $H^{++}$
decay in the IH.  This is shown in Fig.~\ref{Majorana2}.
We see that for $\Phi_1 \approx \pi$ the dominant channels switch to
$e^+ \mu^+,\ e^+ \tau^+$ from $e^+ e^+,\ \mu^+\tau^+$ as in the zero phase limit.
This provides the best hope to probe the Majorana phase.
The decays $H^{\pm} \to e^+_i \bar{\nu}$, on the other hand, are independent of
the unknown Majorana phases, leaving the BR predictions robust. Therefore,
using the lepton violating decays of the singly charged Higgs one can
determinate the neutrino spectrum without any ambiguity.
This is one of our main results of our Letter.

\begin{figure}[tb]
\includegraphics[scale=1,width=5.0cm]{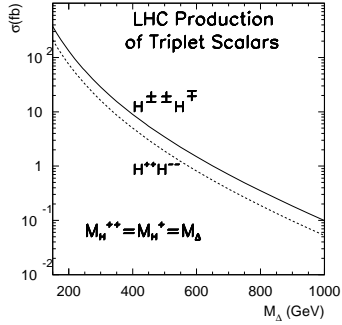}
\caption{\small{Total cross sections in units of fb for $pp\to H^{++} H^{--}$ and $H^{\pm\pm}H^{\mp}$
production versus its mass at $\sqrt s=14$ TeV.}}
\label{total}
\end{figure}

%%%%%%%%%%%%%%%%%%%%%%%%%%%%
\noindent
{\bf{III. Testing the Model at the LHC}}\\
We consider the following production channels
\begin{eqnarray}
\nonumber
q \bar{q} \to \gamma^*,Z^*\to H^{++} H^{--},\ \  \text{and} \ \ q \bar{q'} \to W^* \to  H^{\pm\pm} H^{\mp}.
\end{eqnarray}
The total cross sections versus the mass at the LHC  are shown in Fig.~\ref{total}.
The cross sections range in $100 - 0.1$ fb for a mass of 200$-$1000 GeV, leading to a
potentially observable signal  with a high luminosity.
The associated production $H^{\pm\pm} H^{\mp}$ \cite{Akeroyd:2005gt}
is crucial to test the triplet nature of $H^{\pm\pm}$ and $H^{\pm}$.

\noindent
III.A {\textit{Purely Leptonic Modes}} \\
For $\vd < 10^{-4}$ GeV, we wish to identify as many channels
of leptonic flavor combination as possible to study the
neutrino mass pattern. The $e$'s and $\mu$'s are experimentally
easy to identify, while $\tau$'s can be identified via
their simple charged tracks (1-prong and 3-prongs). We make use of the important
feature that the $\tau$'s from the heavy Higgs decays are highly relativistic
and the missing neutrinos are collimated along the charged tracks, so that
the $\tau$ momentum $p(\tau)$ can be reconstructed effectively.
In fact, we can reconstruct
up to three $\tau$'s if we assume the Higgs pair production with equal masses \cite{big}.
The fully reconstructable signal events are thus
\bea
\nonumber
H^{++} H^{--} &\to&
\ell^+\ell^+\  \ell^-\ell^-, \ \   \ell^\pm \ell^\pm\  \ell^\mp \tau^\mp, \ \
 \ell^\pm \ell^\pm\  \tau^\mp  \tau^\mp,\\
 \nonumber
 &&  \ell^+\tau^+\  \ell^-\tau^-,\ \  \ell^\pm \tau^\pm\  \tau^\mp\tau^\mp, \\
 H^{\pm\pm} H^{\mp} & \to & \ell^\pm\ell^\pm\ \ell^\mp \nu, \ \
 \ell^\pm\ell^\pm\ \tau^\mp \nu,  \nonumber
\eea
where $\ell=e,\mu$. We have performed a full kinematical analysis for those modes,
including judicious cuts to separate the backgrounds, energy-momentum smearing
to simulate the detector effects, and the $p(\tau)$ and $\md$ reconstruction.
We find our kinematical reconstruction procedure highly efficient, with about $50\%$
for $\md=200$ GeV and even higher for a heavier mass.
With a 300 fb$^{-1}$ luminosity, there will still be several reconstructed events in
the leading channels up to $\md\sim$ 1 TeV with negligible backgrounds.

We summarize the leading reconstructable channels and their achievable
branching fractions in Table~\ref{tab:llll}. The $H^\pm$ decays
are robust to determinate the mass pattern since they are independent
of the Majorana phases. For more details see~\cite{big}.
\begin{table}[tb]
\begin{tabular}{| c| c| c|}
\hline
Channels & Modes and BR's(NH) & Modes and BR's (IH)\\
% & Normal Hierarchy & Inverted Hierarchy \\
\hline
$H^{++} H^{--}$ & & \\
$\Phi_1,\Phi_2=0$
& $\mu^+\mu^+ \mu^- \mu^-\ (40\%)^2$  & $e^+e^+ e^-e^-\  (50\%)^2$ \\
& $\mu^+\mu^+ \mu^- \tau^-\   40\%\times 35\% $
&  $e^+e^+ \mu^-\tau^-\  50\%\times 25\%$ \\
& $\mu^+\mu^+ \tau^- \tau^-\   (40\%)^2$  &  $\mu^+\tau^+ \mu^-\tau^-\ (25\%)^2$ \\
& $\mu^+\tau^+  \mu^- \tau^-\   (35\%)^2$  & \\
& $\mu^+\tau^+  \tau^- \tau^- \  35\%\times 40\%$  & \\
$\Phi_1\approx \pi$ & same as  above &
$ee, \mu \tau \to e\mu, e\tau\  (50\%)^2$ \\
$ \Phi_2\approx \pi$ &
$\mu\mu,\tau\tau: \times 1/2, \ \mu \tau: \times 2$ &  same as above  \\
\hline
$H^{\pm\pm} H^{\mp}$ &  &  \\
$\Phi_1,\Phi_2=0$
& $\mu^+\mu^+ \mu^- \nu\   40\%\times 60\%$  &
$e^+e^+ e^- \nu\   (50\%)^2$ \\
& $\mu^+\mu^+  \tau^- \nu \   40\%\times 60\%$  &  \\
$\Phi_1\approx \pi $ & same as  above &
$ee \to e\mu, e\tau\  60\%\times 50\%$ \\
$\Phi_2\approx \pi$ &
$\mu\mu: \times 1/2$  &  same as above \\
\hline
\end{tabular}
\caption{
Leading fully reconstructable leptonic channels and their achievable
branching fractions.}
\label{tab:llll}
\end{table}

\noindent
III.B {\textit{Gauge Boson and Heavy Quark Modes}} \\
For $\vd > 2\times 10^{-4}$ GeV, the dominant decay modes of the heavy Higgs bosons
are the SM gauge bosons. The decay $H^{\pm\pm}\to W^\pm W^\pm$ is governed by $\vd$ and
$H^\pm \to\ W^\pm H_1,\ t\bar b$ by the mixing $\mu$, and $H^\pm \to W^\pm Z$
by a combination of both. Therefore, systematically studying those channels would
provide the evidence of the triplet-doublet mixing and further confirm the seesaw
relation $v_{\Delta}=\mu v_0^2 / \sqrt{2} M_{\Delta}^2$. We have once again
performed detailed signal and background analysis
at the LHC for those channels. We are able to obtain a $20\%$
signal efficiency and a signal-to-background ratio $1:1$ or better.
With a 300 fb$^{-1}$ luminosity, we can achieve statistically significant
signals up to $\md\approx 600$ GeV~\cite{big}.

%%%%%%%%%%%%%%%%%%%%%
\noindent
{\bf{IV. Summary}}\\
The feasibility to test the Type II seesaw mechanism at the LHC has been studied.
We first emphasize the importance to observe the associated production $H^{\pm\pm} H^{\mp}$
to establish the triplet nature of the Higgs field.
In the optimistic scenarios, $v_{\Delta} < 10^{-4}$ GeV, one can test the theory
up to $\md\sim $ 1 TeV, by identifying the leading decay channels as {\it either}
$H^{++} \to \tau^+ \tau^+,\ \mu^+ \mu^+,\ \mu^+ \tau^+$,
$H^{+} \to \tau^+ \bar \nu$ in the Normal Hierarchy, {\it or}
$H^{++} \to e^+ e^+,\ \mu^+ \tau^+$, $H^{+} \to e^+ \bar \nu$ in the Inverted Hierarchy.
If the Majorana phases play an important role, then the $H^{++}$ decay channels are much
less predictable. Always one can use the $H^{\pm}$ decays to determinate the neutrino
spectrum since those are independent of the Majorana phases. For a special case
in the IH, the significant changes in the decay rate of $H^{++}$ with
$e^+ e^+, \mu^+\tau^+ \leftrightarrow e^+ \mu^+, e^+\tau^+$ offer the best hope to probe $\Phi_1$.
In a less favorable scenario, $v_{\Delta} > 2\times 10^{-4}$ GeV, the leptonic channels
are suppressed. The decays $H^{\pm\pm}\to W^\pm W^\pm$ indicate the existence of $\vd$, while
the decays $H^+ \to t \bar{b}$ and $H^+ \to W^+ H_1$ are due to the mixing
between the SM Higgs and the triplet. Statistically significant signals
are achievable up to $\md \approx 600$ GeV. In the most optimistic situation,
$v_{\Delta} \sim10^{-4}$ GeV, the leptonic and
gauge boson channels may be available simultaneously.

\noindent
{\it Acknowledgments:}
{\small We thank E.~Ma and L.-T. Wang for discussions.
This work was supported in part by the U.S.~Department of Energy
under grants DE-FG02-95ER40896, DE-FG02-08ER41531
and the Wisconsin Alumni Research Foundation.
The work at the KITP was supported in
part by the National Science Foundation
under Grant No. PHY05-51164.}
%%%%%%%%%%%%%%%%%%%%%%%%%%%%%

\end{document}